# Demonstration of the Applicability of 3D Slicer to Astronomical Data Using $^{13}$CO and C$^{18}$O Observations of IC 348

Michelle A. Borkin[1], Naomi A. Ridge[2], Alyssa A. Goodman[2], & Michael Halle[3]

[1] Harvard University, Department of Astronomy, 60 Garden St., Cambridge, MA, 02138, USA; mborkin@cfa.harvard.edu

[2] Harvard-Smithsonian Center for Astrophysics, 60 Garden St., Cambridge, MA, 02138, USA

[3] Surgical Planning Lab, Department of Radiology, Brigham and Women's Hospital and Harvard Medical School, 75 Francis St., Boston, MA, 02115, USA

## Abstract

3D Slicer, a brain imaging and visualization computer application developed at Brigham and Women's Hospital's Surgical Planning Lab, was used to visualize the IC 348 star-forming complex in Perseus in RA-DEC-Velocity space. This project is part of a collaboration between Harvard University's Astronomy Department and Medical School, and serves as a proof-of-concept technology demonstration for Harvard's Institute for Innovative Computing (IIC). 3D Slicer is capable of displaying volumes (data cubes), slices in any direction through the volume, 3D models generated from the volume, and 3D models of label maps. The 3D spectral line maps of IC 348 were created using $^{13}$CO and C$^{18}$O data collected at the FCRAO (Five College Radio Astronomy Observatory) by Ridge et al. (2003). The 3D visualization makes the identification of the cloud's inner dense cores and velocity structure components easier than current conventional astronomical methods. It is planned for 3D Slicer to be enhanced with astrophysics-specific features resulting in an astronomical version to be available online through the IIC and in conjunction with the National Virtual Observatory (NVO).

# 1. Introduction

## 1.1 Project Motivation

The motivation behind this project was to apply medical visualization software to astronomical data in order to enhance the understanding and interpretation of the structure of molecular clouds. Preliminary discussions around the formation of the IIC (Harvard Institute for Innovative Computing)[1] brought forth the possibility of using software from the Surgical Planning Lab[2] at Brigham and Women's Hospital called "3D Slicer"[3] to accomplish this goal. Data "cubes" for IC 348 that contain the intensity of a particular molecular transition as a function of line-of-sight velocity (a spectrum), and of each spectrum's two dimensional position on the sky (RA, DEC), were read into 3D Slicer and then 3D surface models, contour maps, and label maps using output from Clumpfind (an algorithm, discussed later in this paper, which makes 3D contours and divides the cloud into clumps) were created. This project demonstrates the possibility and benefits of further adaptation of 3D Slicer for astronomical purposes.

## 1.2 Astrophysical Motivation

The use of visualization tools in the field of star formation is vital to understanding the structure of the cloud and for the location of dense cores or clumps. It is within these dense regions that stars are formed in tight clusters (Lada 1992). Knowing the location of these dense cores along with the location of young stars in a cloud are important in understanding the physical processes of star forming regions. Determining the location and mass of these dense cores shows through their distribution where active star formation is occurring. With properties of these dense cores, specifically the size, shape, and mass spectra, various physical processes including fragmentation can be better studied (Lada, Bally, & Stark 1991). One can also study the initial mass function (IMF), which is a functional relation between the numbers of stars that form per mass interval per unit volume, by comparing the clump mass spectra to the stellar mass spectra within a particular cloud (Lada, Bally, & Stark 1991).

---

[1] http://cfa-www.harvard.edu/~agoodman/IIC/
[2] http://splweb.bwh.harvard.edu:8000/index.html
[3] http://www.slicer.org

The velocity component, essential to determining many of the cloud properties discussed, is hard to visualize and often left out of the analysis. For example, integrated maps take the velocity components along a particular line of sight and combine them to give a single intensity value. These intensity values are usually displayed either in the form of contour outlines or in colors according to an intensity range. Having a way to display the data in a three dimensions in order to take into account RA, DEC, and velocity will make the detection of dense cores, jets, shells, or other structures within the cloud easier and provide a wealth of information compared to a two dimensional display of the data.

## 1.3 Current Astronomical Visualization Tools

There are a variety of astronomical visualization toolkits currently available. These includes programs such as DS9[4], Aipsview[5], and Karma[6]. Specifically looking at their capabilities in displaying 3D cubes, DS9 is able to display a cube as individual slices and play through it as a movie but spectral data cannot be displayed. Aipsview and Karma allow the user to play through the cube as a movie and to examine the spectrum at each point. Karma also does have a feature to display orthogonal slices through a data cube, but this cannot be overlaid with other data or viewed in a three dimensional space. All of these programs lack the ability to display the cube in any 3D form or venue. There are some programs, like IDL[7], that can create 3D views of a data cube, but lack an intuitive user interface. Having a program that can display volume slices and 3D models all in a easy to use interface would go beyond the capabilities in visualization of these programs.

---

[4] http://hea-www.harvard.edu/RD/ds9/
[5] http://aips2.nrao.edu/docs/user/Aipsview/Aipsview.html
[6] http://www.atnf.csiro.au/computing/software/karma/
[7] http://www.rsinc.com/idl/

**1.4 3D Slicer**

3D Slicer excels in offering the user the option of displaying multiple slices in any direction through the original cube, and at the same time displaying multiple models generated from the cube with or without the volume slices. 3D Slicer was originally developed at the MIT Artificial Intelligence Laboratory and the Surgical Planning Lab at Brigham and Women's Hospital. It was designed to help surgeons in image guided surgery (specifically biopsies and craniotomies), to assist in pre-surgical preparation, to be used as a diagnostic tool, and to help in the field of brain research and visualization (Gering 1999). 3D Slicer is built on the ITK and VTK programming toolkits developed as part of the Visible Human Project[8] whose goal is to create full and detailed 3D models of male and female human bodies. ITK and VTK can be used to extend 3D Slicer to include features specifically needed for astrophysical research as will be described later in this paper. A key feature of 3D Slicer is the ability to manipulate, change, and view the data from all angles and orientations in a three dimensional work space.

An example of the main user interface of 3D Slicer is shown in Figure 1. The software was designed to be able to do "registration, segmentation, visualization, and volumetric analysis" (Gering 1999). In other words, be able to automate the alignment of images (registration), define regions of different intensity or density, visualize in three dimensions, and make quantitative measurements. The software is a freely distributed product and is in use at multiple research institutes and hospitals.

---

[8] http://www.nlm.nih.gov/research/visible/visible_human.html

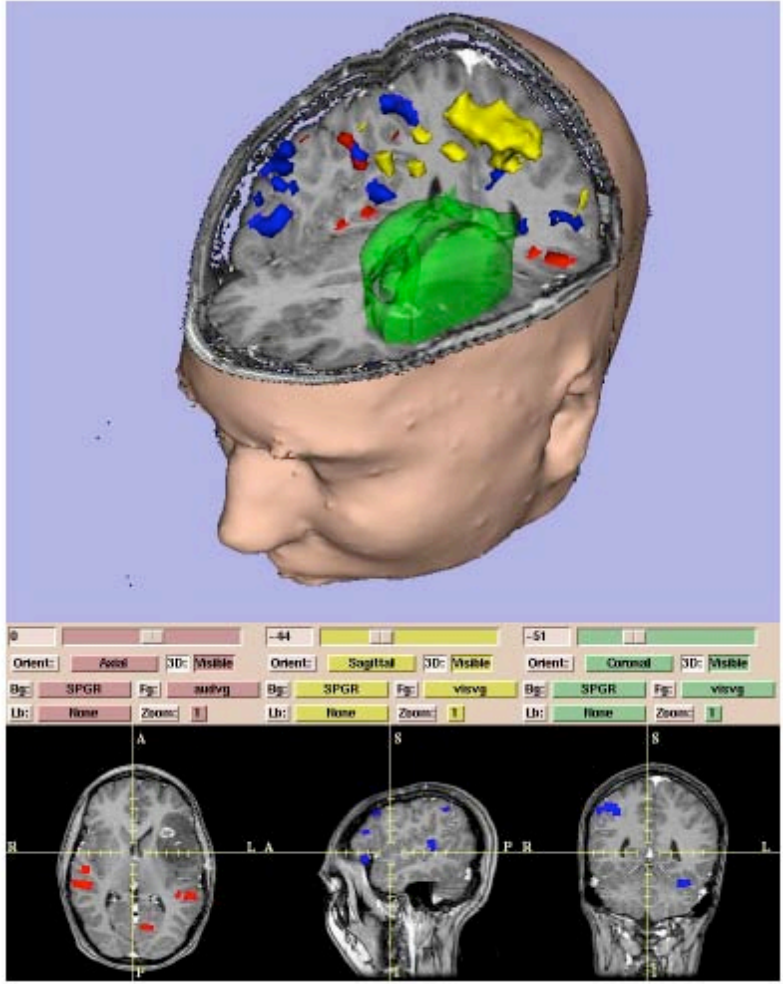

**Figure 1:** The use of 3D Slicer to detect a tumor (green) in an MRI scan of a brain. (From Gering, et al. 1999)

## 1.5  IC 348

IC 348 is a nearby young star forming cluster located in the Perseus molecular cloud (see Figure 2).  Using the Hipparcos parallax data, it is at a distance of approximately 260±25 pc (Cohen, Herbst, & Williams 2004).  Its age is in the range of 1.3 and 3 million years old based on the average of the pre-main sequence stars (D'Antona & Mazzitelli 1994).  It has a number of outflows including HH 211 first observed by McCaughrean et al. (1994) and has been studied in the visual (Trullols & Jordi 1997, Herbig 1998), near-infrared (Lada & Lada 1995, Luhman et al. 1998, 2003, 2005), and x-ray (Preibisch, Zinnecker, & Herbig 1996, Preibisch & Zinnecker 2002) wavelengths from which hundreds of cluster members have been identified (Cohen, Herbst, & Williams 2004).

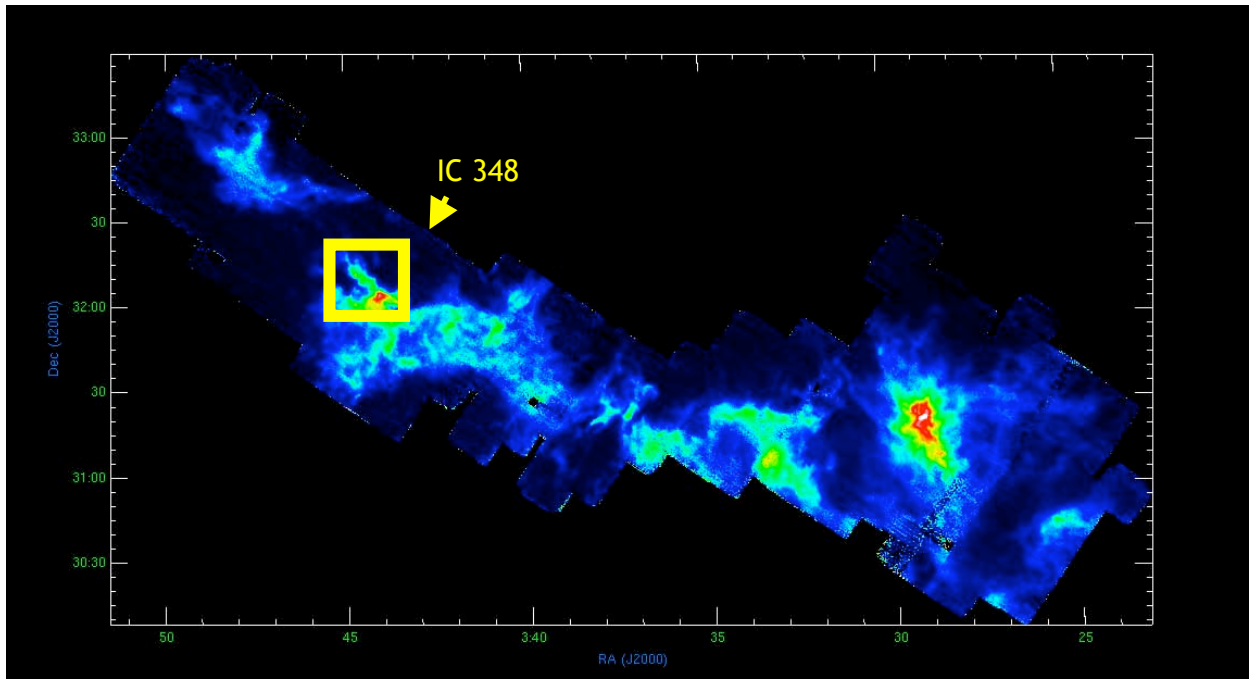

**Figure 2:** Location of IC 348 on an integrated intensity map of $^{13}$CO emission in Perseus.

IC 348 is a good source to study since it has a lot of active star formation, has been well observed, and is relatively compact. For this project, IC 348 was analyzed in $^{13}$CO and C$^{18}$O to study the gaseous structure of the cloud. Viewing the star forming region with various molecular tracers is advantageous since they each show different regions or properties of a cloud. In the case of $^{13}$CO and C$^{18}$O, the $^{13}$CO will show the whole cloud whereas the C$^{18}$O will only display the densest regions. This is due to the fact that a certain density of CO is required in order to be detected. At low densities, C$^{18}$O is not seen since it is not abundant enough to be detected, but at high column densities $^{13}$CO (and eventually C$^{18}$O) will become optically thick so it cannot trace the true column density of the structure. Both $^{13}$CO and C$^{18}$O are optically thick at the densities inside cores (>$10^4$ cm$^{-3}$), but the sample being analyzed is on a spatial scale around 1 pc and the average densities are approximately 100-1000 cm$^{-3}$ so we do not have to worry about this. Displaying the information for both sets of data, however, can create a more complete picture of the cloud.

## 2. Data

The data used here was collected in 2002 at the 14 meter FCRAO (Five College Radio Astronomy Observatory) telescope with the SEQUOIA focal plane array.[9] The receiver was used with a digital correlator providing a total bandwidth of 25 MHz over 1024 channels. The J = 1 – 0 transition of $^{13}$CO and the J = 1 – 0 transition of C$^{18}$O were observed simultaneously using an on-the-fly mapping technique (OTF). Both maps are 30 x 30 arcminutes in size and have a velocity resolution of 0.07 km s$^{-1}$ (Ridge, et al. 2003). Using IDL, a series of 300 RA-DEC map slices, each 73 x 73 pixels in size, were made representing the velocity range of 0 to 20 km s$^{-1}$ where each of these maps is a single 0.07 km s$^{-1}$ velocity channel. These maps were then converted into 16 bit grayscale maps in order to be read into 3D Slicer. In 3D Slicer, a series of models were created to demonstrate the smoothing and noise reduction capabilities of the program, and a series of models representing 3D contours were created for each of the $^{13}$CO and C$^{18}$O cubes. Figure 3 displays a screen shot of 3D Slicer displaying volume slices and a 3D contour map of IC 348 in $^{13}$CO. (See the Appendix for detailed information on the settings for each specific model.)

3D Slicer also has the capability of reading in label maps (a particular region is represented by a single pixel value). Over the summer of 2004, data cubes were produced by applying the Clumpfind algorithm (Williams, de Geus, & Blitz 1994) to the same $^{13}$CO data in order to locate dense molecular condensations in IC 348 (Borkin, et al. 2004). Clumpfind works by making a contour map through the cube to locate the peak intensities, and then steps down the contour levels to determine where the clumps are located. Clumpfind allows the user to adjust various parameters including minimum clump size, contour spacing, and lowest contour level. For this data, a minimum contour level of 1 K and a contour interval of 0.5 K were set giving an output of 91 clumps. The output data cube was sliced in the same way as those described earlier and read into 3D Slicer as a label map. Each clump was then made into an individual 3D model.

---

[9] This data set of IC 348 in $^{13}$CO and C$^{18}$O from Ridge et al. 2003 was used instead of data from the COMPLETE Survey because the data sets from COMPLETE were too large to complete a detailed analysis on for the timescale of this project and, when this project was started, higher density tracers were not available from COMPLETE.

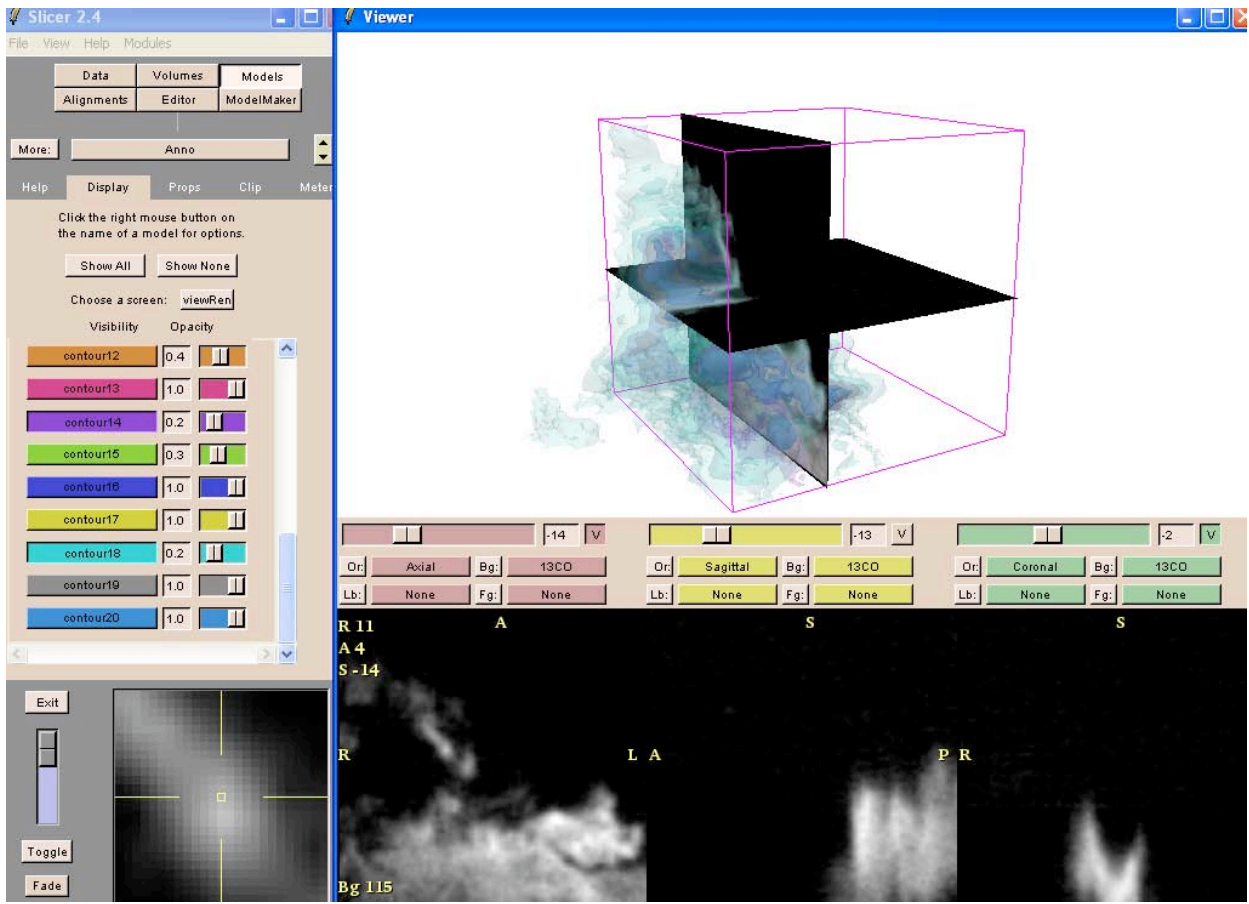

**Figure 3:** Screen shot of 3D Slicer while displaying a 3D contour map of IC 348 in $^{13}$CO. The slices through the model are RA-DEC and RA-Velocity slices which are also displayed along with a DEC-Velocity slice on the bottom toolbar.

## 3. Analysis and Results

The first set of models, displayed in Figure 4, demonstrate the noise reduction and smoothing capabilities of 3D Slicer. Each of these models represent the surface (or outer most 3D contour) of IC 348 based on different criteria. All of the models are made by setting a threshold limit (a minimum intensity) that the model is made from, and the threshold is displayed as a transparently shaded-in region on the volume. The first model is the default threshold limit that 3D Slicer sets which does cut out the noise, but also cuts out too much of the fainter emission around the edges of the cloud. The second model has a manually set threshold which includes all of the cloud, but, in order to get all of the cloud's emission, high intensity noise is also detected and included in the model. The third model has the manually set threshold, but the noise has been removed with the "remove island" function. The remove island function will subtract any regions that are below a set minimum pixel size, but it will not cut the region off if it grows into a larger region in other slices. This model is the optimal version of the surface of IC 348. The fourth model has the manually set threshold and the remove island function applied, but also the "dilation" function. The dilation function cuts a set number of pixels off of the threshold regions, and in this model one pixel was removed around all the edges. This did smooth the model, but it removed too much detail. These models demonstrate 3D Slicer's ability to reduce noise, smooth models, and allow the user to use their discretion while examining the volumes as how to customize each model.

For both the $^{13}$CO and C$^{18}$O data cubes, 3D contour maps were made by creating a series of models from threshold limits in which the limits are equally spaced apart. For the outermost contour levels, the remove island function was applied to remove unwanted noise. In examining Figure 5, the C$^{18}$O surface fits perfectly inside the $^{13}$CO surface which is expected since C$^{18}$O traces the highest density regions due to its lower abundance. In Figures 6 and 7, the inner structures of IC 348 are shown in both molecules. Distinct hierarchical clumping is shown with the higher intensity clumps appearing first in the main part of the cloud (lower right portion of the images) and a larger number of clumps but lower intensity appearing together in the northern region (upper left portion of the images). Figure 8 displays a K-band image from 2MASS, which reveals otherwise dust-enshrouded young forming stars, of IC 348 overlaid with integrated intensity contours of $^{13}CO$. In general this is the standard way to display the information in a data cube and the basis of star to gas comparison. Even though the contours highlight the main part of the cloud (lower right) as being the peak to focus on, the other peak in the northern portion of the cloud is more important to star formation. As displayed in Figure 9, this same northern region of the cloud indicated by the yellow circle, where the dense clumps are visible in 3D Slicer, corresponds to the cluster of young stars visible in the K band image. Without 3D Slicer, one might also think that the northern and main parts of the cloud were physically connected as it appears in the integrated map (discussed further with Figure 10).

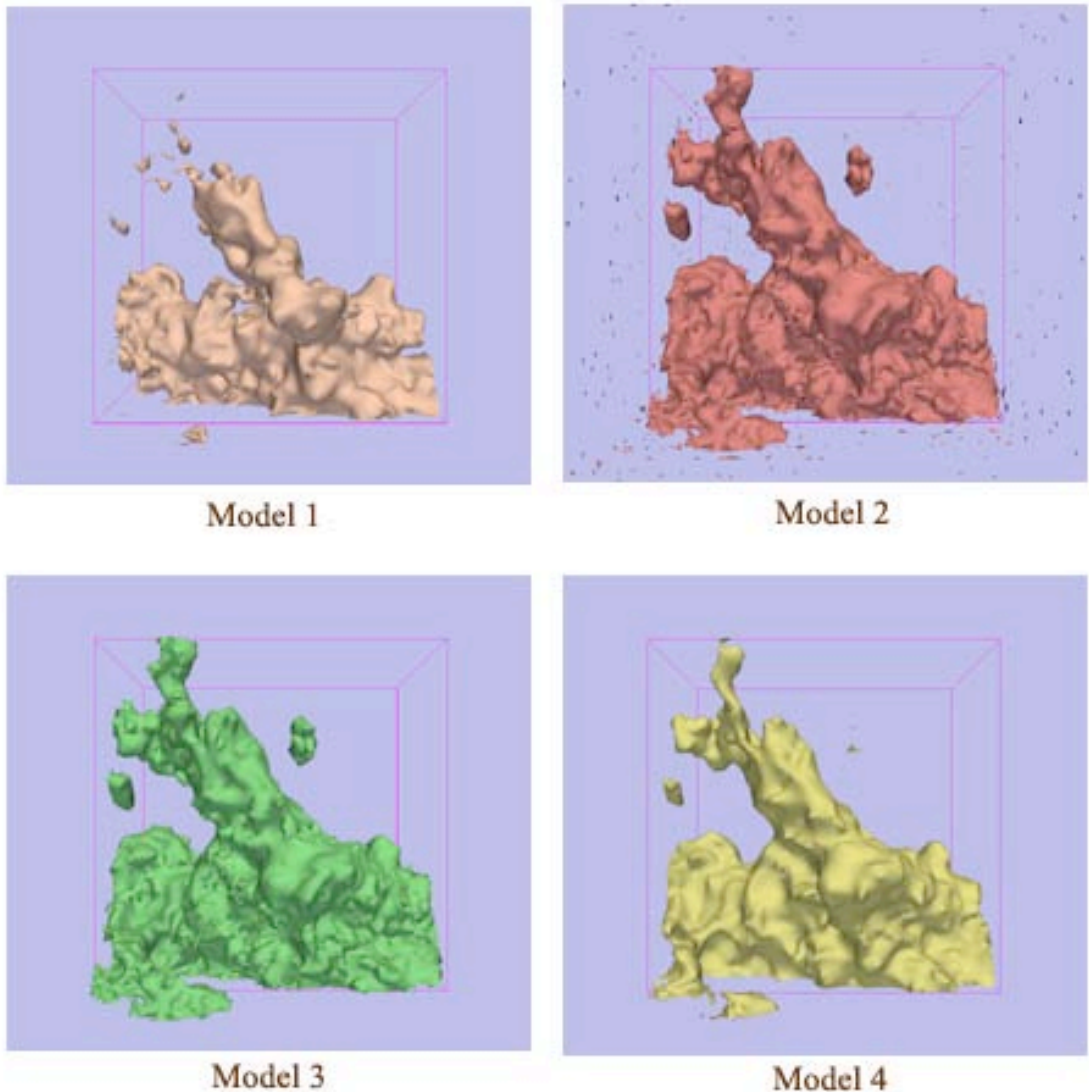

**Figure 4:** Models of the surface of IC 348 in $^{13}$CO using the default threshold settings (Model 1), a manually set threshold (Model 2), a manually set threshold with the remove island function applied (Model 3), and a manually set threshold with the remove island and dilation functions applied (Model 4).

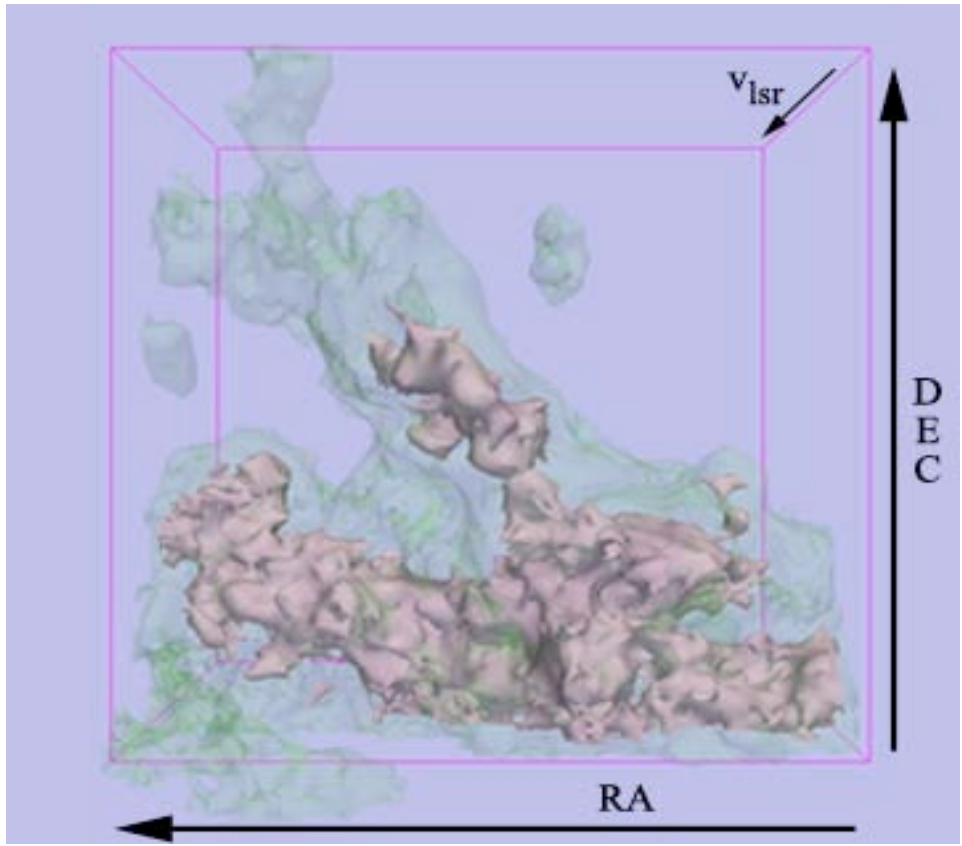

**Figure 5:** The surface of IC 348 in $C^{18}O$ (pink) and $^{13}CO$ (green).

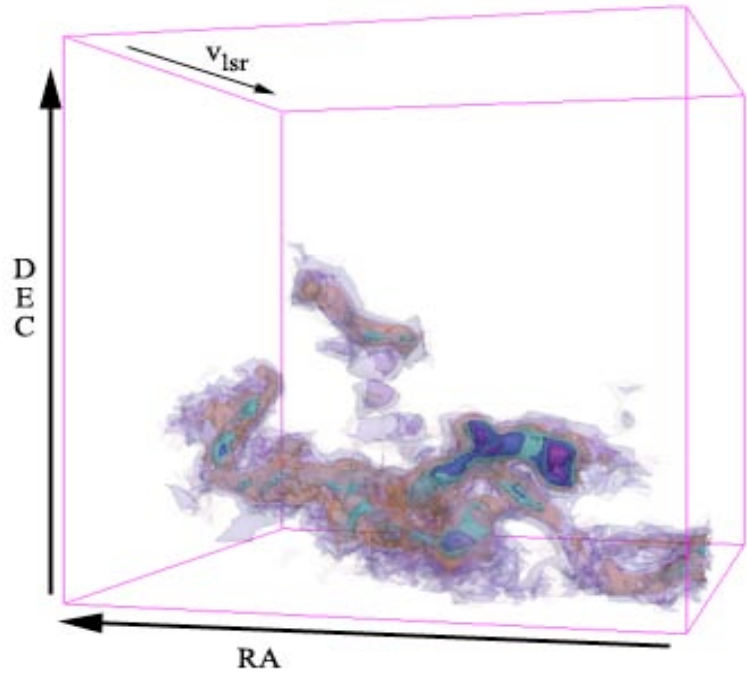

**Figure 6:** 3D contours of IC 348 in $C^{18}O$.

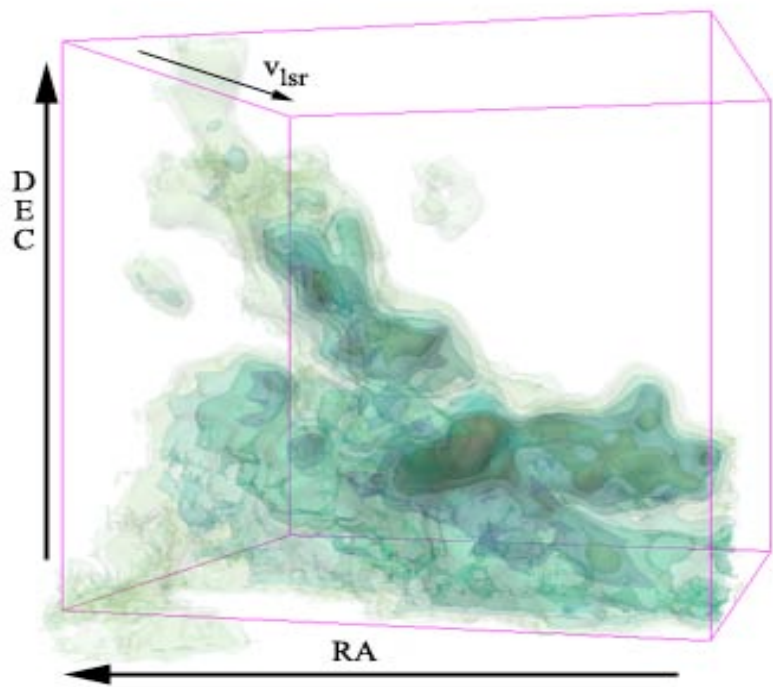

**Figure 7:** 3D contours of IC 348 in $^{13}CO$.

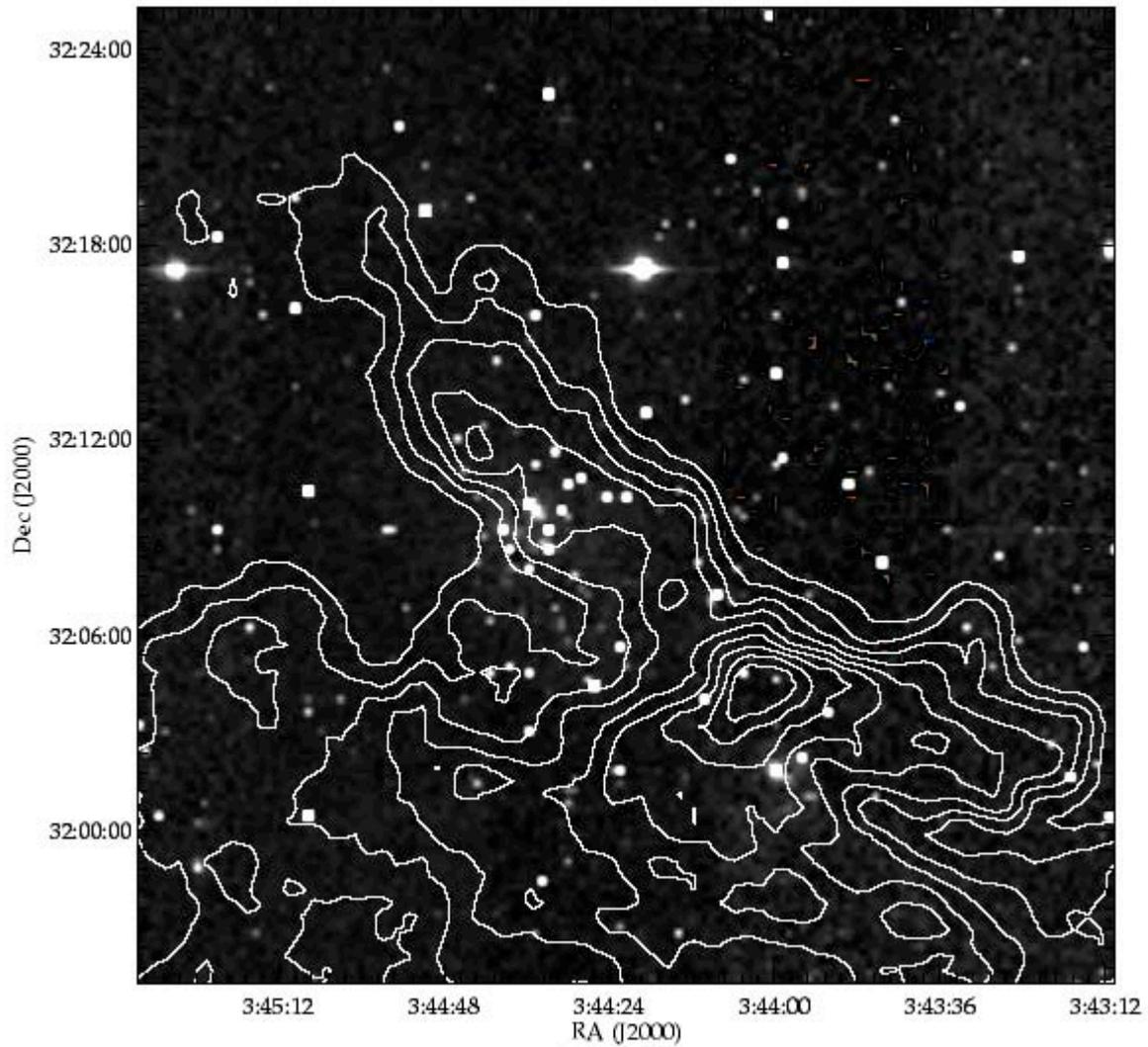

**Figure 8:** K-band image of IC 348 with an overlay of $^{13}$CO integrated intensity contours. The contours go from 10-100% of maximum intensity in 10% intervals.

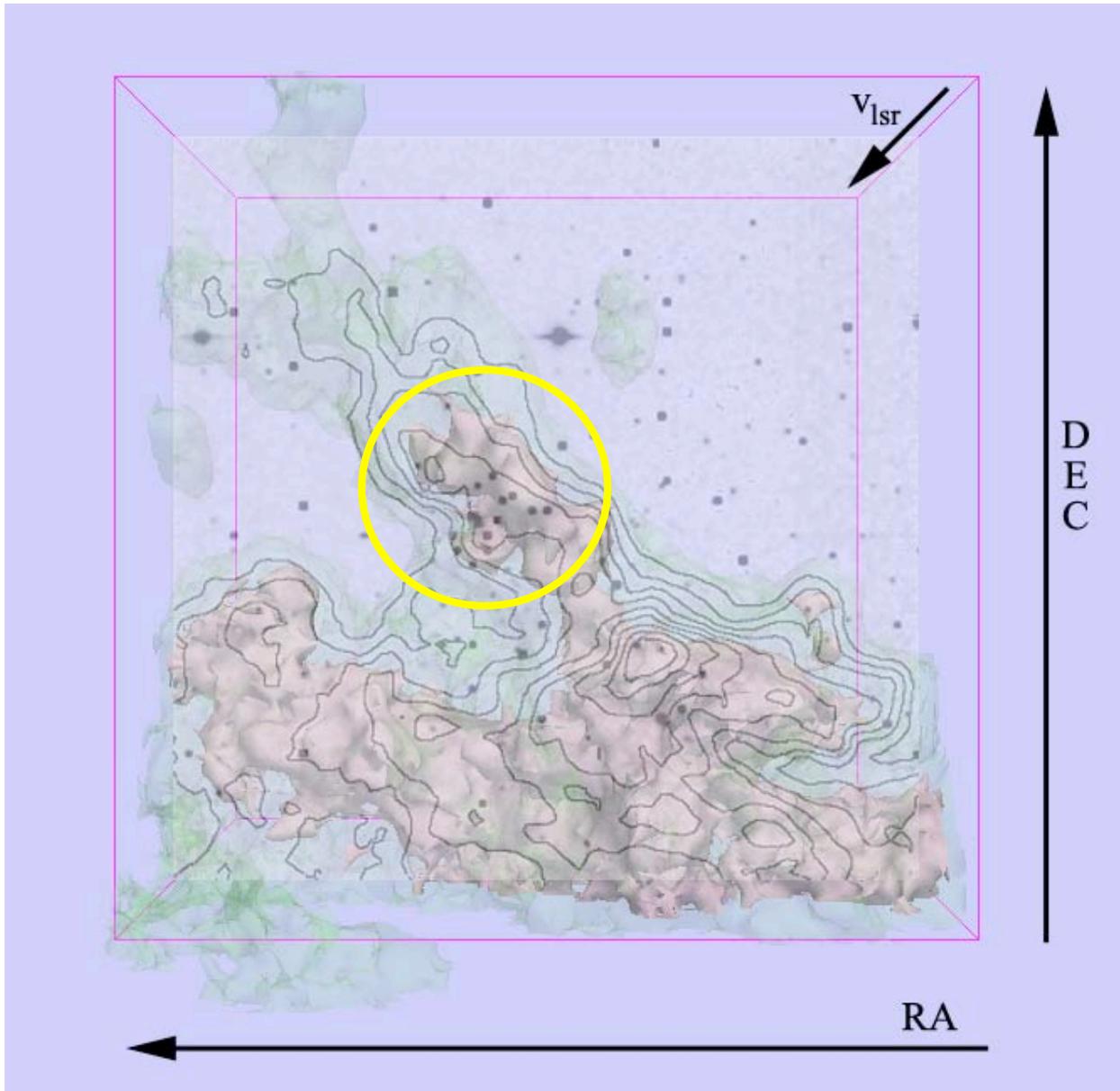

**Figure 9:** Overlay of a color inverted version of Figure 8 ($^{13}$CO integrated contours on a K band image of IC 348) on Figure 5 (surface of IC 348 in C$^{18}$O (pink) and $^{13}$CO (green)). The yellow circle highlights a young cluster of stars that match a prominent structure displayed in 3D Slicer, but which is not obviously correlated with the integrated contours.

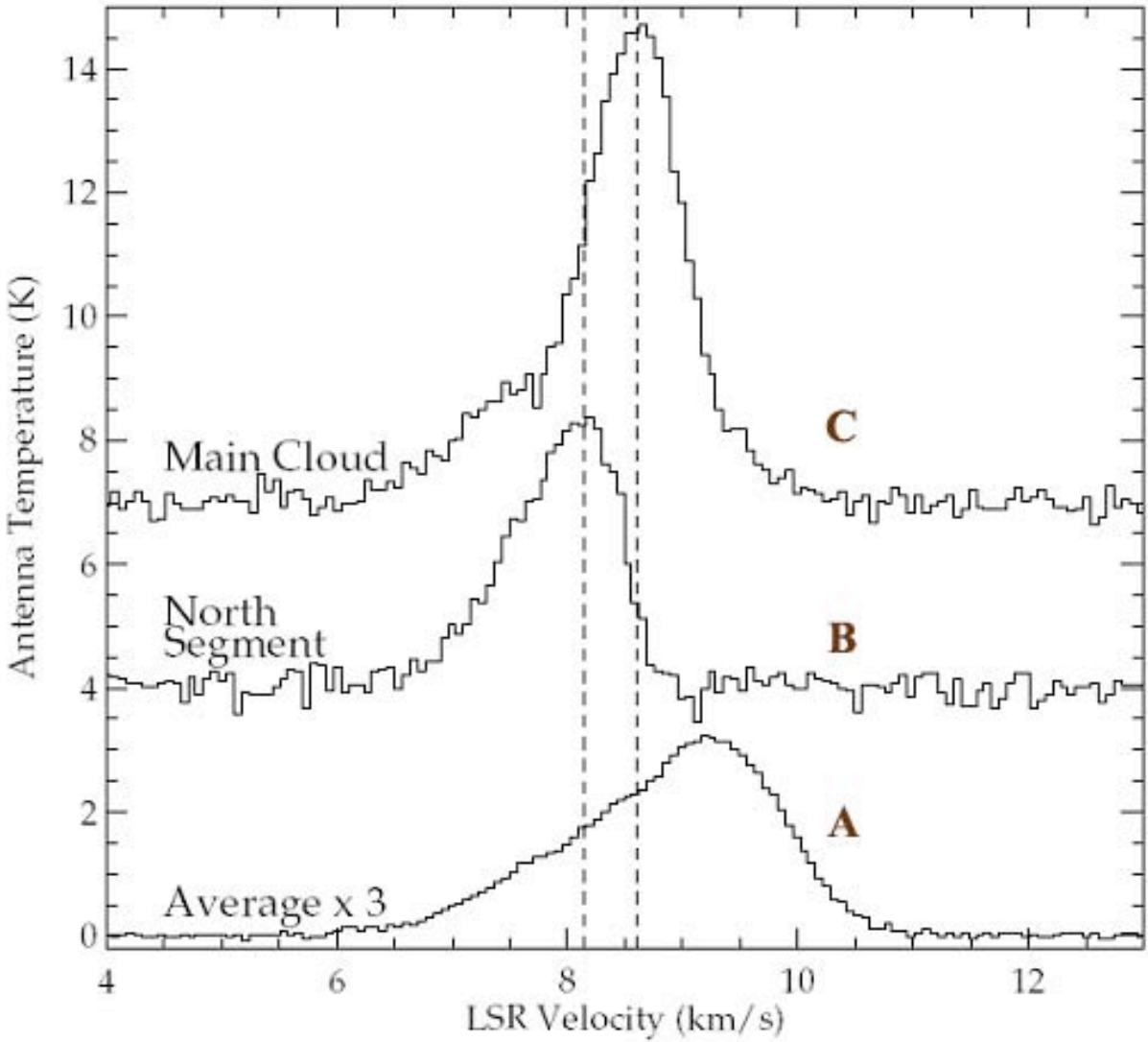

**Figure 10:** Plotted is the average spectrum multiplied by three (A), the spectrum for the brightest point of the north portion (B), and the spectrum for the brightest point of the main portion of IC 348 (C). The peak for B is at 8.15 km s$^{-1}$ and the peak for C is at 8.6 km s$^{-1}$.

3D Slicer is also able to display distinct regions of the cloud that have differing velocities. From a separate analysis of IC 348 data, Figure 10 shows (going from bottom to top) the average spectrum for the cloud (spectrum A), the spectrum for the brightest part of the north (upper left) portion of IC 348 (spectrum B), and the spectrum for the brightest part of the main (lower right) portion of IC 348 (spectrum C). One can see that the peak for spectrum B (8.15 km s$^{-1}$) is at a slightly lower velocity than that of spectrum C (8.6 km s$^{-1}$). This slight difference in velocity is distinctly visible in 3D Slicer as demonstrated in Figure 11. In these models the northern region of cloud (spectrum B) is clearly separated in velocity space from the rest of the cloud (spectrum C) confirming the implications derived from the spectra. This would have been hard to detect with existing astronomical applications.

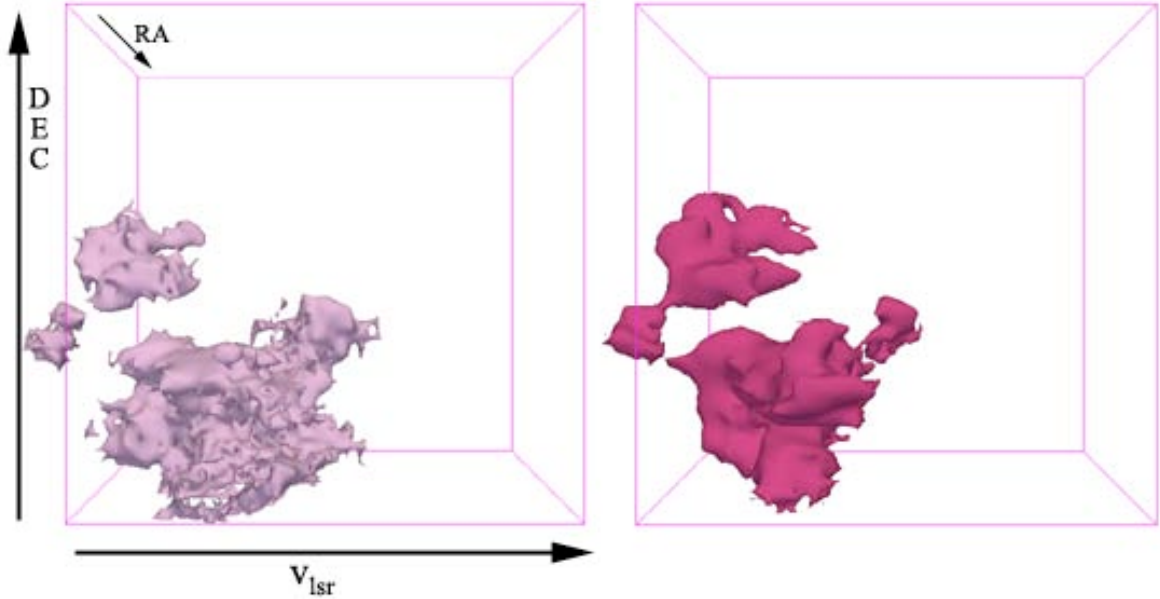

**Figure 11:** Velocity-DEC plots for the surface in $C^{18}O$ and an inner contour in $^{13}CO$ of IC 348.

The third set of models is made from the Clumpfind data as shown in Figure 12. Clumpfind goes through each channel map in the data cube, divides it into clumps via a contouring algorithm in which the user enters the minimum contour level, follows each clump through the cube and prints out a label map where the clumps are assigned a number going from highest to lowest intensity. Clumpfind does divide the cloud into clumps, but it does not generate a hierarchical structure with smaller clumps embedded in larger ones. Figure 12 shows in the left column select channel maps from IC 348 data cube with purple representing the lowest intensity to red at the highest intensities. In the right column are the corresponding Clumpfind label maps for each particular channel map in which each clump is randomly assigned a color. The appearance of the clumps as adjacent volumes stuck to each other (as seen in Figure 13.a.) is emblematic of the limitations of Clumpfind – no clump can be inside another, and there are no gaps between clumps.

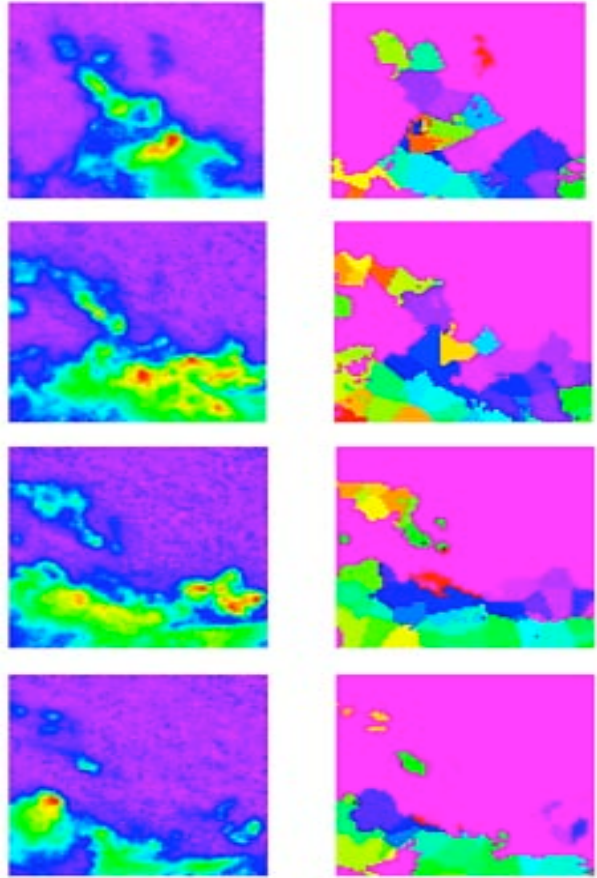

**Figure 12:** Displayed are channel maps of IC348 where the original $^{13}$CO emission data in the left column with the corresponding Clumpfind image in the right column. Velocities going from top to bottom: 4.12, 4.65, 5.18, and 5.58 km s$^{-1}$.

    The label maps created by Clumpfind of the $^{13}$CO clumps for IC 348 were read into 3D Slicer, and then a threshold model was made for each individual clump. Figure 13.a. shows a

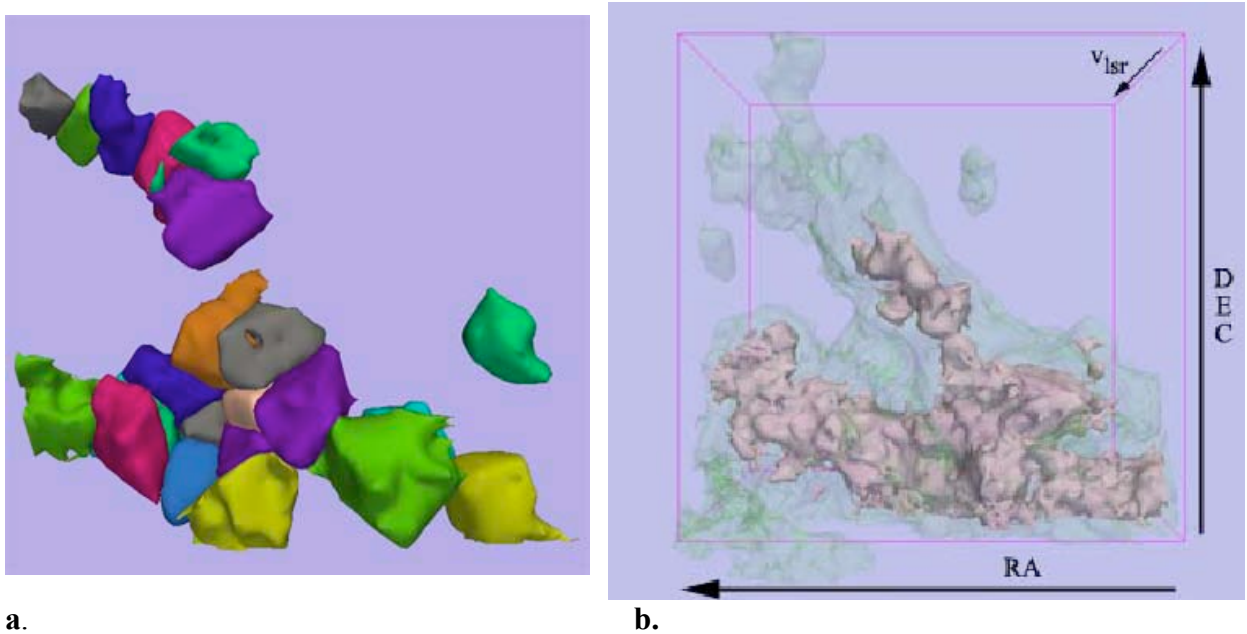

**a.**                                                 **b.**

**Figure 13:** Side by side comparisons of: **a.** select clump models, where each color represents a distinct clump, generated from the $^{13}$CO Clumpfind output for IC 348 and **b.** surfaces of IC 348 in C$^{18}$O (pink) and $^{13}$CO (green) as displayed in 3D Slicer (same as Figure 5).

select number of clumps displaying the distinct northern branch of IC 348 and the main portion of the cloud. As compared to the 3D contour maps created in 3D Slicer, Figure 13.b., which show the distinct hierarchical and nesting appearance of gaseous clumps, the Clumpfind models look crude in comparison and beg the question of how accurate the Clumpfind algorithm is in describing the cloud. The results of calculating the number of clumps, the number of self-gravitating clumps, or the size of clumps will be very different with the 3D contour maps created in 3D Slicer versus Clumpfind.

      These clump models can be overlaid with the other 3D contours or the Clumpfind volume and specific threshold limits can be moved through the other generated models. Having features such as the label maps, generation of 3D models, volume display and overlay, and the ability to move through and around the models in three dimensional space makes the analysis of the cloud easier than in conventional visualization packages.

## 4. Conclusion

This project demonstrates the application of 3D Slicer to astronomical data and the program's potential for future use in analyzing data cubes. 3D contour maps were created for IC 348 in $^{13}$CO and C$^{18}$O which display the portions of the cloud moving at different velocities and where there is inner clumping. Models were also created using Clumpfind label maps created for the $^{13}$CO data. Future additions to 3D Slicer to make it more useful for astronomical data analysis would include a FITS file reader, the ability to display coordinate information both along the volume border and the 3D cube border, the ability to place labels anywhere in the viewer space not just on the model surfaces, the ability to handle data sets smaller than 70 x 70 pixels, and the addition of a custom color palette (to include colors with more useful names other than "Brain" or "Tumor"). This new astronomical version of 3D Slicer will be available online through the IIC and in conjunction with the National Virtual Observatory[10].

Using the present version of 3D Slicer, and the enhanced version once it is ready, more star forming regions will be analyzed. Specifically, data gathered as part of the COMPLETE[11] (CoOrdinated Molecular Probe Line Extinction Thermal Emission) Survey of the Perseus, Ophiuchus, and Serpens star forming regions will be analyzed and visualized in the same manner as demonstrated in this paper, including the comparison of various molecular tracers, wavelengths, and Clumpfind label maps. In order to complete this analysis, the use of more powerful computers is needed given the size of the data sets (more than 150,000 spectra in each compared to around 5,000 in the IC 348 data set).

In addition to conducting the same type of analysis as demonstrated in this paper, other tools in 3D Slicer and other applications of the program will be used. It is possible to measure the sizes of clumps by either using the measuring tool or by reading out the polygon dimensions from the 3D Slicer data files of the generated 3D models. 3D Slicer also has built in ITK "segmentation" tools developed to find specific "features" (i.e. parts of the brain) based on properties of the image. These segmentation tools can be used to identify clumps and generate lists containing their location, size, and mass. By determining these characteristics for all the clumps in a cloud, the mass spectra for the cloud can be calculated and the detailed analysis of the cloud's properties can be studied including the IMF. In addition to overlaying three dimensional surface contours, the addition of overlaying plots will be another way to identify and compare features in the models. For example, an image or plot of young stars will be converted to have the proper scale, pixel dimensions, and projection so that it could be moved through the model as a volume. Another application of 3D Slicer to be tested is the comparison of theoretical models to actual clouds. Both real and simulated spectral-line cubes can be looked at and statistics gathered on clumps. 3D Slicer is a useful tool at present and will be an even better tool in the near future to study the physical processes that occur within a star forming cloud dictating its evolution and star producing properties.

---

[10] http://cfa-www.harvard.edu/nvo/
[11] http://cfa-www.harvard.edu/COMPLETE/

**Acknowledgments**

We would like to thank Marianna Jakab and the staff of the SPL at Brigham and Women's Hospital for all their time and help in instruction, computer assistance, and valuable insight into both the practical usage and creative vision in the application of 3D Slicer. This research was supported in part by the National Alliance for Medical Image Computing (NIH BISTI grant 1-U54-EB005149) and the Neuroimage Analysis Center (NIH NCRR grant 5-P41-RR13218).

## 6. Appendix

Tables 1 and 2 are a complete list of all the models created for this paper to test the noise reduction capabilities of 3D Slicer, and to create 3D contour maps for IC 348 in $^{13}$CO and C$^{18}$O. The columns, going from left to right, give the file name of the model, the intensity range for the threshold, the minimum island size in pixels if the function was applied, and the number of pixels dilated if the function was applied.

| $^{13}$CO Models | Threshold minimum (pixel intensity); maximum = 255 | Remove Island size (pixels) | Dilation (pixels) |
|---|---|---|---|
| SmoothedSurface | 15 | 10 | 1 |
| 13COSurface | 15 | 10 | - |
| default | 51 | - | - |
| noisy | 15 | - | - |
| contour1 | 225 | - | - |
| contour2 | 215 | - | - |
| contour3 | 205 | - | - |
| contour4 | 195 | - | - |
| contour5 | 185 | - | - |
| contour6 | 175 | - | - |
| contour7 | 165 | - | - |
| contour8 | 155 | - | - |
| contour9 | 145 | - | - |
| contour10 | 135 | - | - |
| contour11 | 125 | - | - |
| contour12 | 115 | - | - |
| contour13 | 105 | - | - |
| contour14 | 95 | - | - |
| contour15 | 85 | - | - |
| contour16 | 75 | - | - |
| contour17 | 65 | - | - |
| contour18 | 55 | - | - |
| contour19 | 45 | - | - |
| contour20 | 35 | - | - |
| contour21 | 25 | 10 | - |

**Table 1:** List of model names and settings created for IC 348 in $^{13}$CO.

| C$^{18}$O Models | | |
| --- | --- | --- |
| | Threshold minimum (pixel intensity) | Remove Island size (pixels) |
| c18oSurface | 38 | 10 |
| C18ocontour1 | 205 | - |
| C18ocontour2 | 195 | - |
| C18ocontour3 | 185 | - |
| C18ocontour4 | 175 | - |
| C18ocontour5 | 165 | - |
| C18ocontour6 | 155 | - |
| C18ocontour7 | 145 | - |
| C18ocontour8 | 135 | - |
| C18ocontour9 | 125 | - |
| C18ocontour10 | 115 | - |
| C18ocontour11 | 105 | - |
| C18ocontour12 | 95 | - |
| C18ocontour13 | 85 | - |
| C18ocontour14 | 75 | - |
| C18ocontour15 | 65 | 10 |
| C18ocontour16 | 55 | 10 |
| C18ocontour17 | 45 | 10 |

**Table 2:** List of model names and settings created for IC 348 in C$^{18}$O.